\documentclass[conference]{IEEEtran}
\IEEEoverridecommandlockouts
\usepackage{cite}
\usepackage{amsmath,amssymb,amsfonts}
\usepackage{graphicx}
\usepackage{textcomp}
\usepackage{xcolor}
\usepackage{colortbl}
\usepackage[colorlinks=true,
            linkcolor=red,
            urlcolor=blue,
            citecolor=green!50!black!80!]{hyperref}
\usepackage{graphicx}%
\usepackage{multirow}%
\usepackage{amsmath,amssymb,amsfonts}%
\usepackage{amsthm}%
\usepackage{mathrsfs}%
\usepackage{xcolor}%
\usepackage{textcomp}%
\usepackage{manyfoot}%
\usepackage{booktabs}%
\usepackage{algpseudocode}%
\usepackage{listings}%
\usepackage{makecell}
\usepackage{enumitem}
\usepackage{url}
\usepackage{algorithm}
\usepackage{algpseudocode}
\usepackage{amsmath}
\usepackage{pifont}
\newcommand{\cmark}{\ding{51}}%
\newcommand{\xmark}{\ding{55}}%

\usepackage{amsmath,amsfonts,bm}

\def\eqref#1{equation~\ref{#1}}

\def\1{\bm{1}}

\def\vb{{\bm{b}}}

\def\vh{{\bm{h}}}

\def\mE{{\bm{E}}}

\def\mH{{\bm{H}}}

\def\mN{{\bm{N}}}

\def\mV{{\bm{V}}}
\def\mW{{\bm{W}}}

\DeclareMathAlphabet{\mathsfit}{\encodingdefault}{\sfdefault}{m}{sl}
\SetMathAlphabet{\mathsfit}{bold}{\encodingdefault}{\sfdefault}{bx}{n}

\def\gE{{\mathcal{E}}}

\def\gG{{\mathcal{G}}}

\def\gP{{\mathcal{P}}}

\def\gV{{\mathcal{V}}}

\def\sL{{\mathbb{L}}}

\newcommand{\R}{\mathbb{R}}

\newcommand{\eg}{\textit{e}.\textit{g}., }
\newcommand{\ie}{\textit{i}.\textit{e}., }

\newcommand{\loca}{\textsc{ProtLOCA}}

\newcounter{bxincomm}
\definecolor{aqua}{rgb}{0.00,0.67,0.80}

\newcounter{lrcomm}
\definecolor{limegreen}{rgb}{0.2,0.7,0.2}

\newcounter{tycomm}
\definecolor{orange}{rgb}{0.2,0.7,0.2}

\newcounter{todocomm}

\def\BibTeX{{\rm B\kern-.05em{\sc i\kern-.025em b}\kern-.08em
    T\kern-.1667em\lower.7ex\hbox{E}\kern-.125emX}}
\begin{document}

\title{Protein Representation Learning with\\Sequence Information Embedding:\\Does it Always Lead to a Better Performance?\\
\thanks{This work was supported by the National Natural Science Foundation of China (11974239; 62302291), the Innovation Program of Shanghai Municipal Education Commission (2019-01-07-00-02-E00076), Shanghai Jiao Tong University Scientific and Technological Innovation Funds (21X010200843), the Student Innovation Center at Shanghai Jiao Tong University, and Shanghai Artificial Intelligence Laboratory.}
}

\makeatletter
\newcommand{\linebreakand}{%
  \end{@IEEEauthorhalign}
  \hfill\mbox{}\par
  \mbox{}\hfill\begin{@IEEEauthorhalign}
}
\makeatother

\author{
\IEEEauthorblockN{Yang Tan}
\IEEEauthorblockA{
\textit{Shanghai Jiao Tong University}\\
Shanghai, China \\
tyang@mail.ecust.edu.cn}
\and 
\IEEEauthorblockN{Lirong Zheng}
\IEEEauthorblockA{
\textit{University of Michigan}\\
MI, USA \\
lrzheng@umich.edu}
\and
\IEEEauthorblockN{Bozitao Zhong}
\IEEEauthorblockA{
\textit{Shanghai Jiao Tong University}\\
Shanghai, China \\
zbztzhz@gmail.com}
\linebreakand
\IEEEauthorblockN{Liang Hong}
\IEEEauthorblockA{
\textit{Shanghai Jiao Tong University}\\
Shanghai, China \\
hong3liang@sjtu.edu.cn}
\and
\IEEEauthorblockN{Bingxin Zhou}
\IEEEauthorblockA{
\textit{Shanghai Jiao Tong University}\\
Shanghai, China \\
bingxin.zhou@sjtu.edu.cn}
}

\maketitle

\begin{abstract}
Deep learning has become a crucial tool in studying proteins. While the significance of modeling protein structure has been discussed extensively in the literature, amino acid types are typically included in the input as a default operation for many inference tasks. This study demonstrates with structure alignment task that embedding amino acid types in some cases may not help a deep learning model learn better representation. To this end, we propose \loca, a local geometry alignment method based solely on amino acid structure representation. The effectiveness of \loca~is examined by a global structure-matching task on protein pairs with an independent test dataset based on CATH labels. Our method outperforms existing sequence- and structure-based representation learning methods by more quickly and accurately matching structurally consistent protein domains. Furthermore, in local structure pairing tasks, \loca~for the first time provides a valid solution to highlight common local structures among proteins with different overall structures but the same function. This suggests a new possibility for using deep learning methods to analyze protein structure to infer function.
\end{abstract}


\section{Introduction}
In recent years, an increasing number of studies have designed deep learning-based solutions to understand the construction principles of proteins, including tasks like structure folding \cite{senior2020improved}, sequence design \cite{madani2023large}, and function prediction \cite{yu2023enzyme}. These attempts are based on the important relationship deduced by biologists that protein sequence determines structure and structure determines function \cite{koehler2023sequence}. However, the complex composition and numerous variables of protein molecules make this relationship extremely intricate, preventing the creation of a simple theoretical system. Additionally, experimental validation is too costly to test all possible proteins. Therefore, deep learning algorithms have been designed to help discover from large databases the mapping relationships between protein sequence, structure, and function. An increasing number of studies have developed deep learning methods to solve specific biological problems, achieving great success in validation across various downstream tasks \cite{sapoval2022current}.

The dominant methods for protein representation learning currently focus on feature extraction from protein sequences, due to the abundance of amino acid sequence data and the development of language models. On the other hand, with the development of structure prediction models \cite{jumper2021highly,abramson2024accurate}, protein structure datasets have also become significantly enriched, leading some studies to utilize geometric deep learning methods \cite{jing2020gvp,zhou2023protein, zhang2022gearnet} to extract three-dimensional protein structures and incorporate them into hidden representations. Moreover, recent studies have found that incorporating both sequence and structure information can further enhance the expressivity of the embeddings, leading to better performance in prediction tasks such as mutation effect prediction \cite{tan2023protssn, wang2022lm-gvp} and binding-affinity prediction \cite{li2021structure_binding}.

\begin{figure*}[th]
    \centering
    \includegraphics[width=\textwidth]{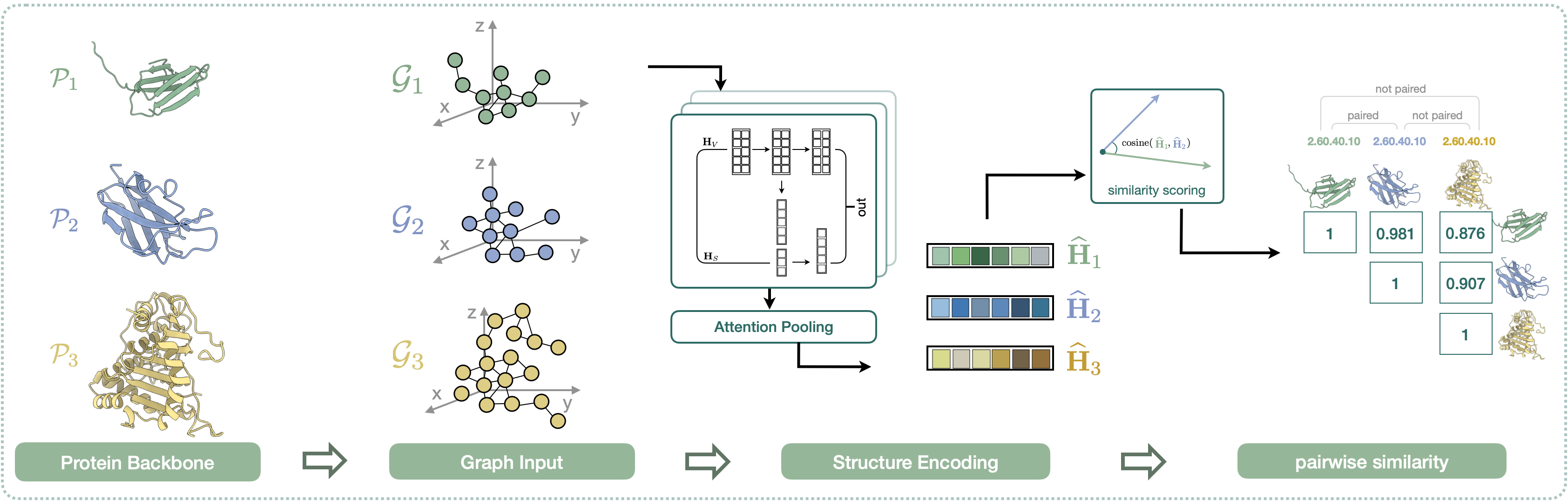}
    \caption{An illustrative pipeline of \loca~for structure pairing (see Section~\ref{sec:globalMatch}). We employ \loca~to extract protein vector representations for protein structures and calculate the cosine similarity between the learned hidden representation of protein pairs. }
    \label{fig:architectureGlobal}
\end{figure*}

Nowadays, unless performing sequence inference (\ie sequence data is not used as input), amino acid sequence information is always included in the model input. Unlike the necessity of structure embedding, which has been discussed extensively by various studies \cite{hsu2022esmif,yang2022mifst,notin2023proteingym}, to the best of our knowledge, no work has explored explicitly the role of incorporating sequence information into neural networks in any form. This leads us to propose the following research question: 
\textit{\textbf{Is sequence information really always a beneficial element for any protein representation learning task?}}
To address this question, we delve into the structure alignment task, where the inference objective is to determine the similarity between two protein structures. We compare the performance of models trained with and without amino acid sequence information. As shown in Tables~\ref{tab:result_main} and Fig.~\ref{fig:sensitivity}, we find that including sequence information interferes with the prediction in the structure matching task. This observation aligns with biological intuition: highly dissimilar sequences can still fold into similar structures. Therefore, incorporating sequence features when summarizing protein structural characteristics may dilute the important information in the learned embeddings, leading to significant matching errors. On the other hand, although sequences generally determine protein structures, in some special cases, highly similar protein sequences can fold into different structures. Hence, matching structures based on sequence information may introduce additional errors.

While sequence information is not always beneficial for certain inference tasks on proteins, such as local structure alignment, it is natural to ask: \textit{\textbf{how to encode structural information effectively for amino acids for sequence-irrelevant tasks?}} This paper introduces \loca~for {\underline{PROT}}ein {\underline{LOC}}al structure {\underline{A}}lignment. The model processes the three-dimensional structure of the protein with roto-equivariant graph neural networks to extract vector representations of the amino acid local geometry. The proposed \loca~is validated on two tasks of protein structure alignment. In global protein structures matching (Fig.~\ref{fig:architectureGlobal}), we assign binary classification labels for protein domains, where the ground-truth label is defined by the CATH classification system \cite{sillitoe2021cath}. \loca~achieves state-of-the-art performance over various sequence-based and structure-based protein feature extraction methods. For the second task of local structure alignment (Fig.~\ref{fig:architectureLocal}), we leverage \loca~to find common local folding in proteins that have different overall structures. We select a crucial type of regulator in gene processes called DNA binding protein, whose local structure for DNA regulation shares a similar fold while their overall structures differ \cite{takeda1983dna}. Among these two DNA binding proteins from different species, \loca~effectively identified the common local structure that is crucial for its function, while the overall structures between them are different. In comparison, existing global alignment methods like TM-align \cite{zhang2005tm} fail to locate such local similarity.

In summary, this study contributes in three aspects.
\begin{enumerate}[leftmargin=*]
    \item We find that amino acid sequence information is not always beneficial for encoding effective representations for protein inference tasks and demonstrate through an important structural biology task of structure alignment. 
    \item We separate an independent subset from CATH4.3 and introduce \textbf{CATH-aligns} and \textbf{CATH-aligns+}, two standard structure matching benchmark datasets based on high-quality protein domain labels. We also provide a comprehensive comparison of popular sequence-based and structure-based protein encoding methods on the two benchmarks.
    \item We propose \loca, a protein structure embedding method that achieves state-of-the-art performance on global protein structure alignment tasks. Additionally, we validate \loca~on a specific task, demonstrating its effectiveness in identifying similar local structures.
\end{enumerate}

\section{Global Structure Matching}
\label{sec:globalMatch}
\subsection{Problem Formulation}
Consider three arbitrary peptide chains $\gP_1$, $\gP_2$, and $\gP_3$, where $\gP_1$ and $\gP_2$ share a similar global structure. While $\gP_3$ has a significantly different overall structure, it contains a common substructure $\gP^{\prime}$ with $\gP_1$ and $\gP_2$. A \emph{global structure matching} evaluates the overall similarity of peptide chain pairs, \eg assigns a high similarity score to the $(\gP_1, \gP_2)$ pair and a low similarity score to both $(\gP_1, \gP_3)$ and $(\gP_2, \gP_3)$.

\begin{figure*}[th]
    \centering
    \includegraphics[width=\textwidth]{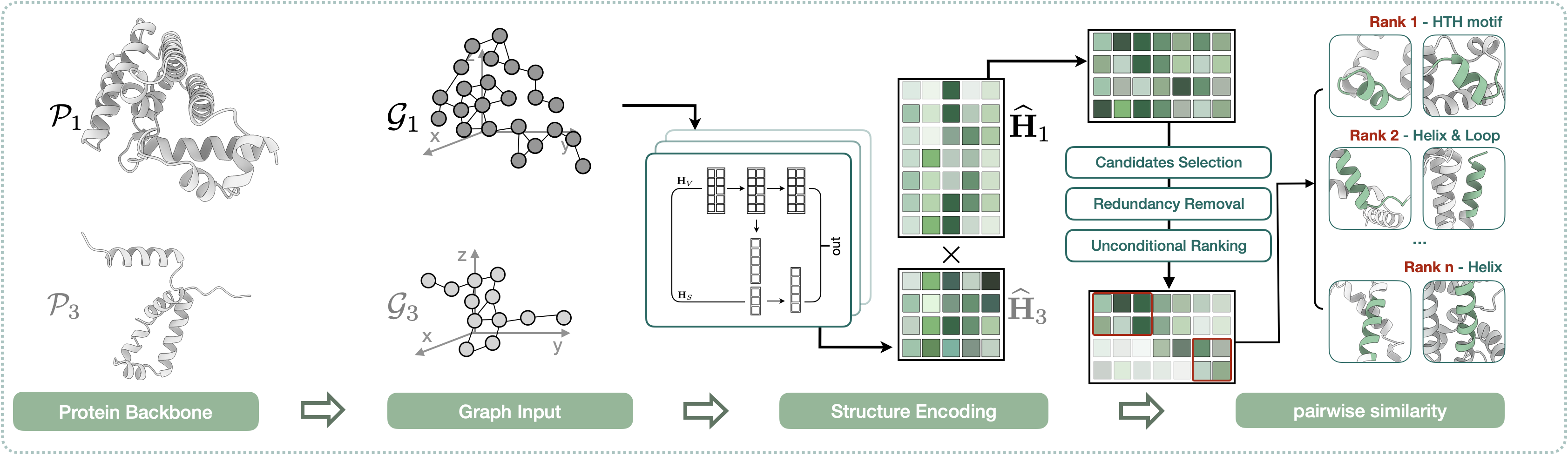}
    \caption{An illustrative pipeline of \loca~for local structure alignment (see Section~\ref{sec:localMatch}). We employ \loca~for residue-level point-to-point matching, which identifies similar local structures on proteins with different overall structures. }
    \label{fig:architectureLocal}
\end{figure*}

\subsection{Feature Representation}
Define $\gG = (\gV, \gE)$ the graph representation of a peptide chain's backbone. Each node $v\in\gV$ represents an amino acid, and spatially closed nodes (\ie Euclidean distance smaller than 10\AA) are connected by directed edges $e\in\gE$. For the $i$th amino acid, the node feature is composed of scalars and vectors, \ie $\mH_v^{i} = (\mN_S^i, \mN_V^i)$. The scalar feature $\mN_S^i$ contains one-hot encodings of structure tokens, such as DSSP-based secondary structure \cite{kabsch1983dssp} or FoldSeek embedding \cite{van2024fast}. The vector feature $\mN_V^i\in\R^{3\times 3}$ summarizes the spatial relationship of neighborhood heavy atoms along the sequence, including two directional vectors by the coordinates of the $C_{\alpha}$ atoms ($\mN_{V,1}^i=C_{\alpha_{i+1}}-C_{\alpha_i}$; $\mN_{V,2}^i=C_{\alpha_{i-1}}-C_{\alpha_i}$) and a tetrahedral geometry unit vector
\begin{equation*}
    \mN_{V,3}^i=\sqrt{\frac13}\frac{(\mathbf{n}\times\mathbf{c})}{\|\mathbf{n}\times\mathbf{c}\|_2} - \sqrt{\frac23}\frac{(\mathbf{n}+\mathbf{c})}{\|\mathbf{n}+\mathbf{c}\|_2},
\end{equation*}
where $\mathbf{n}=N_i-C_{\alpha_i}$ and $\mathbf{c}=C_i-C_{\alpha_i}$.

Similarly, on the edge of two connected nodes from $v_i$ to $v_j$, we define edge features $\mH_e^{ij}$ by scalar features and vector features. The scalar feature $\mE_S^{ij}\in\R^{32}$ concatenates the radial basis functions (RBF) representations \footnote{We use 16 Gaussian radial basis functions with centers evenly spaced between 0 and 20\AA.} of $\|C_{\alpha_j}-C_{\alpha_i}\|_2$ and sinusoidal positional encoding \footnote{We use the positional encoding method described in Transformer \cite{vaswani2017transformer}.} of the relative Euclidean distance between $v_i$ and $v_j$. The vector feature $\mE_V^{ij}\in\R^3$ is defined by the direction of $C_{\alpha_{i}}-C_{\alpha_j}$.

\subsection{Model Architecture}
\loca~implements geometric vector perceptrons (GVP) \cite{jing2020gvp} to extract scalar and vector features from the nodes and edges of protein graphs. For an arbitrary protein graph $\gG$, a \emph{GVP layer} computes embeddings for the scalar feature $\mH_S$ and the vector feature $\mH_V$, \ie
\begin{equation}
\label{eq:gvp}
    (\mH_S^{\prime}, \mH_V^{\prime}) = {\rm GVP}(\mH_S, \mH_V).
\end{equation}
The key to a $\rm GVP(\cdot)$ layer is composed of multiple iterations of \emph{scalar-vector propagations}, defined in (\ref{eq:gvp}). At the $(\ell+1)$th ($0\leq\ell< L$) iteration,
\begin{equation}
\label{eq:sv_prop}
\begin{aligned}
    \mH_S^{(\ell+1)}&= \sigma\big(\mW_1\cdot{\rm concat}({\rm norm}(\mW_2\mH_V^{(\ell)}),\mH_3)+\vb\big), \\
    \mH_V^{(\ell+1)}&= \sigma\big({\rm norm}(\mV^{(\ell+1)})\big)\odot\mV^{(\ell+1)}, \\
    \text{where } \mV^{(\ell+1)}&=\mW_{4}\mW_2\mH_V^{(\ell)}.
\end{aligned}
\end{equation}
Here $\mW_1, \mW_1, \mW_3, \mW_4$ and $\vb$ are learnable parameters for this layer, $\odot$ denote row-wise multiplication, ${\rm norm}(\cdot)$ denotes row-wise $\sL_2$ normalization, and $\sigma(\cdot)$ represents the sigmoid activation function. At $\ell=0$, the initial input $(\mH_S^{(0)}, \mH_V^{(0)}) = (\mH_S, \mH_V)$. At the last layer when $\ell+1=L$, it outputs $(\mH_S^{\prime}, \mH_V^{\prime}) = (\mH_S^{(L)}, \mH_V^{(L)})$. We set $L=3$ in each of the $\rm GVP(\cdot)$ layers.

The separately encoded scalar and vector representations $(\mH_S^{\prime}, \mH_V^{\prime})$, before sending to further prediction, are combined to obtain an AA-level matrix representation. 
This requires additional transformations, which we define as a \emph{GVP Transform layer}. 
As introduced below, we first define a concatenated feature $\mH={\rm concat}(\mH_S^{\prime}, \mH_V^{\prime})$. For the $i$th node and the edge of connected nodes $i\rightarrow j$, we define:
\begin{equation}
\label{eq:conv}
\begin{aligned}
    \mathbf{h}_{m}^{ij} &:=
    {\rm GVP}\left(\mathrm{concat}\left(\mathbf{h}_{\mathbf{v}}^{j}, \mathbf{h}_{e}^{ij}\right)\right) \\
    \mathbf{h}_{\mathbf{v}}^{i} &\leftarrow 
    \mathrm{LayerNorm}\left(\mathbf{h}_{\mathbf{v}}^{i} + \frac{1}{k} \mathrm{Dropout}\left(\sum_{j: \mathbf{e}_{ij} \in \mathcal{E}} \mathbf{h}_{m}^{ij}\right)\right), 
\end{aligned}
\end{equation}
where each feature vector $\mathbf{h}$ is a concatenation of scalar features $\mH_S^{\prime}$ and vector features $\mH_V^{\prime}$, 
$k$ is the number of incoming messages from $i$'s neighbors, and $\vh_v^{i}$ and $\vh_m^{ij}$ are the embedding of scalars and vectors for node $i$ and edge $i\to j$. 

We also add an extra \emph{feed-forward layer} when updating the node representation
\begin{equation}
\label{eq:ffn}
    \mathbf{h_v^{i}}\leftarrow\text{LayerNorm}\left(\mathbf{h_v^{i}}+\text{Dropout}\left({\rm GVP}^{\prime}\left(\mathbf{h_v^{i}}\right)\right)\right),
\end{equation}
where ${\rm GVP}^{\prime}$ denotes a GVP layer with $L=2$. We use superscripts to distinguish it from the previous GVP layers in (\ref{eq:gvp}) and (\ref{eq:conv}), which includes three layers of scalar-vector propagations defined in (\ref{eq:sv_prop}). In comparison, ${\rm GVP}^{\prime}$ in (\ref{eq:ffn}) only applies 2 layers of scalar-vector propagation. 

The stack of GVP convolution and feed-forward transformation defined in (\ref{eq:gvp})-(\ref{eq:ffn}) constructs a \emph{GVP-GNN block}. The block is repeated multiple times to obtain expressive node representations. The feed-forward layer is applied at the end of every GVP-GNN block except for the last block. In implementation, we set the reputation to $6$. See ablation studies in Section~\ref{sec:experiment}) for more details.

The node representation is sent to readout layers for label prediction. In the training phase, a dense layer is employed to recovery the input tokens:
\begin{align}
    y = \mathbf{W}(\text{ReLU}(\text{DropOut}(\mathbf{W}\mathbf{H}_S))).
\end{align}
For prediction tasks, \ie global structure matching, we obtain vector representation for the input protein with an normalized average pooling layer, \ie 
\begin{align}
\label{eq:AAembedding}
    \mathbf{h_v} = \big\|\frac{1}{n} \sum_{i=1}^n \mathbf{h_v^{i}}\big\|.
\end{align}

To measure the similarity of a protein pair $(\gP_1,\gP_2)$ with the respective learned vector representations $(\vh_1,\vh_2)$, we define the cosine similarity:
\begin{equation}
\label{eq:cos_sim}
\begin{aligned}
    {\rm sim}(\gP_1,\gP_2)=\frac{\vh_1\cdot\vh_2}{\|\vh_1\|\cdot\|\vh_2\|}.
\end{aligned}
\end{equation}

\subsection{Training Objective}
\label{sec:global_train_objective}
Training \loca~only involves the scalar and vector features extracted from backbone coordinates, excluding inputs directly related to amino acid types. The model is trained in a self-supervised learning manner with the objective of denoising the perturbed node features. Two types of corruption approaches are considered for adding noise, including masking and permutation. In the former, values in $\mN_S$ are set to $0$ with a probability $p$; in the latter, values in $\mN_S$ are randomly replaced by another $\mN_S$ value with a probability of $1-p$. Additional discussion on the tunable parameter $p$ can be found in Section~\ref{sec:experiment}.

\section{Local Structure Alignment}
\label{sec:localMatch}
\subsection{Problem Formulation}
Consider two arbitrary peptide chains $\gP_1$ and $\gP_3$ with different sequence length and overall structure and a common substructure $\gP^{\prime}$. An \emph{local structure alignment} task aims to identify highly similar local regions $\gP^{\prime}$ in the input data $(\gP_1, \gP_3)$. 

\subsection{\loca~for Local Structure Alignment}
In the global structure matching task, we employ average pooling on the amino acid representations of the protein pairs to obtain vector representations for comparing protein-level similarity. However, this simplified method cannot provide insights into the alignment of protein local regions. While functionally similar proteins may only have similar active regions and differ in overall structure, discovering local alignments of proteins could be essential for functional region identification and analysis. To this end, we introduce a modified \loca~with a simple heuristic algorithm to highlight similar regions for protein pairs. After extracting the hidden representation for nodes by (\ref{eq:ffn}), we conduct the following three steps for local alignment identification.

\subsubsection{Candidate Selection}
For two proteins $\gP_1$ and $\gP_3$ with $m$ and $n$ amino acids, respectively, \loca~extracts $256$-dimensional representations $\mH_1\in\R^{m\times 256}$ and $\mH_3\in\R^{n\times 256}$. Similar to the global matching task, we score the similarity between the two matrices by the cosine similarity:
\begin{equation}
    {\rm sim}(\gP_1,\gP_3)= \frac{\mH_1\cdot\mH_3}{\|\mH_1\|\cdot\|\mH_3\|}
\end{equation}
We will use the output similarity matrix ${\rm sim}(\gP_1,\gP_3)\in\R^{m\times n}$ for identifying structurally aligned regions between the two proteins. Intuitively speaking, the similarity scores on the diagonal indicates the point-to-point alignment of the two proteins. By selecting high values on the diagonal, the corresponding structurally aligned local regions of the two proteins are recognized. 

\subsubsection{Redundancy Removal}
To further investigating the regional similarity of the two proteins, we set a similarity threshold $\mu$ for the diagonal and a minimum structure size $s$ for the similar local structure of interest. We first iterate over all possible subset blocks $\mH^{\prime} \in \mH$ along the diagonal line with the size from  $s\times s$ until $m\times n$ along the diagonal line. The possible $\mH^{\prime}$ are those that ${\rm mean}({\rm diag}(\mH^{\prime}))>\mu$. We record the mean and variance for all candidate $\mH^{\prime}$s. The second step removes redundant blocks from the candidates group with an overlap threshold $d$. We traverse all candidate $\mH^{\prime}$. For two arbitrary $\mH^{\prime}_1,\mH^{\prime}_3$, if more than $d$ rows or columns are overlapped, the smaller block matrix will be dropped. After the two steps, we obtain a set of non-overlapping candidate regions. In this study, we set $\mu=10$, $s=0.8$, and $d=5$.


\subsubsection{Unconditional Ranking}
To identify the best matching local structures, we sort the obtained candidates by their variance (calculated from the first step of candidate selection) in ascending order. This unconditional ranking approach assumes no prior knowledge about the specific region to be matched (\eg active site). In cases where the target region is known, we can optionally employ a conditional ranking method. This method sorts the candidates based on the degree of index-level overlap between the query structure and the candidate structures in descending order.

\begin{table*}[!t]
\caption{Performance comparison of baseline models on \textbf{CATH-aligns} for structure alignment. The classification performance is evaluated by AUC. Both the average AUC and the detailed fold-wise AUC are reported. }
\label{tab:result_main}
\begin{center}
\resizebox{\linewidth}{!}{
\begin{tabular}{llcccc>{\columncolor[gray]{0.9}}cccc>{\columncolor[gray]{0.9}}cccc}
    \toprule
    \multicolumn{4}{c}{\textbf{Model Information}} & \multicolumn{2}{c}{\textbf{Input}} & \multicolumn{4}{c}{\textbf{CATH-aligns}}& \multicolumn{4}{c}{\textbf{CATH-aligns+}}\\\cmidrule(lr){1-4}\cmidrule(lr){5-6}\cmidrule(lr){7-10}\cmidrule(lr){11-14}
    \rowcolor{white}
    Type & Name & Version & \# Params & AA & Structure & average & fold 1 & fold 2 & fold 3 & average & fold 1 & fold 2 & fold 3 \\
    \midrule
    \multirow{2}{*}{Aligment} & \multirow{2}{*}{FoldSeek \cite{van2024fast}}  & 3Di & -& \xmark & \cmark & 0.900 & 0.903 & 0.901 & 0.897 & 0.891 & 0.893 & 0.892 & 0.888 \\
    &  & 3Di-AA & -& \cmark & \cmark & 0.888 & 0.889 & 0.888 & 0.886 & 0.881 & 0.882 & 0.881 & 0.879 \\
    \midrule
    \multirow{11}{*}{Embedding} & \multirow{3}{*}{ESM2 \cite{lin2023esm2}} & t33\_650M & 650M& \cmark & \xmark & 0.685 & 0.685 & 0.684 & 0.687 & 0.672 & 0.672 & 0.674 & 0.671 \\
    &  & t36\_3B & 3,000M & \cmark & \xmark & 0.700 & 0.697 & 0.699 & 0.704 & 0.685 & 0.685 & 0.687 & 0.682 \\
    &  & t48\_15B & 15,000M& \cmark & \xmark  & 0.814 & 0.813 & 0.814 & 0.814 & 0.788 & 0.788 & 0.790 & 0.786 \\ \cmidrule(lr){2-14}
    & \multirow{2}{*}{ProstT5 \cite{heinzinger2023prostt5}}  & AA2fold & 3,000M & \cmark & \xmark & 0.907 & 0.905 & 0.909 & 0.908 & 0.851 & 0.851 & 0.852 & 0.850 \\
    &  & fold2AA & 3,000M & \xmark & \cmark & 0.921 & 0.921 & 0.92 & 0.922 & 0.838 & 0.841 & 0.839 & 0.834 \\\cmidrule(lr){2-14}
    & ESM-IF \cite{hsu2022esmif}  & - & 148M & \cmark & \cmark & 0.625 & 0.624 & 0.625 & 0.627 & 0.851 & 0.853 & 0.851 & 0.849 \\ \cmidrule(lr){2-14}
    & MIF-ST \cite{yang2022mifst}  & - &643M& \cmark & \cmark & 0.882 & 0.897 & 0.873 & 0.877 & 0.614 & 0.611 & 0.616 & 0.616 \\
    \cmidrule(lr){2-14}
    & ProtLOCA (Ours) & - &5.9M& \xmark & \cmark & \textbf{0.965} & \textbf{0.966} & \textbf{0.964} & \textbf{0.964} & \textbf{0.895} & \textbf{0.895} & \textbf{0.895} & \textbf{0.895} \\
    \bottomrule
\end{tabular}
}
\end{center}
\end{table*}

\begin{figure*}[ht]
    \centering
    \includegraphics[width=\linewidth]{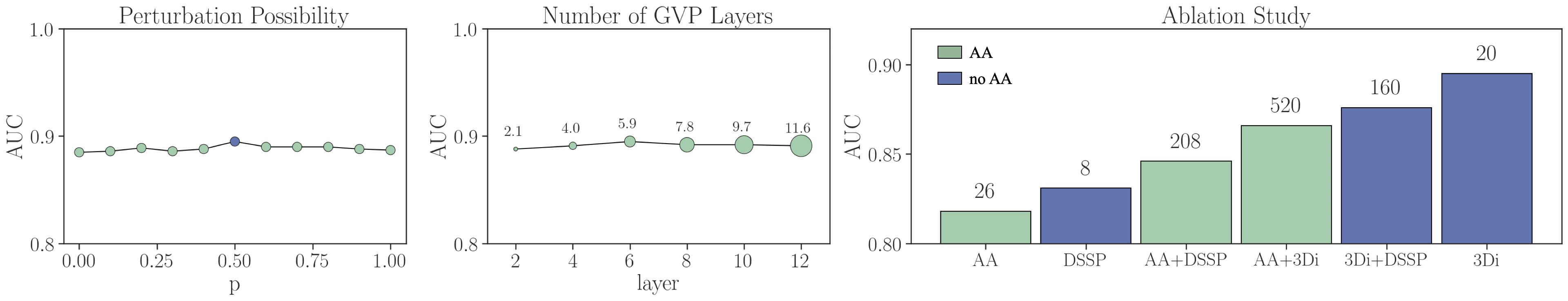}
    \vspace{-5mm}
    \caption{Model performance on different (left) perturbation possibility $p$ on mask corruption; (middle) number of GVP layers; (right) pre-training targets.}
    \label{fig:sensitivity}
\end{figure*}

\section{Experimental Analysis}
\label{sec:experiment}
\loca~is pre-trained on an unlabeled protein structure dataset from \textbf{CATH4.3} (introduced below). We examine \loca~on protein structure alignment tasks involving both global structure matching and local structure alignment. For the global structure matching task, we provide quantitative comparisons with baseline methods on two independent benchmark datasets, \textbf{CATH-aligns} and \textbf{CATH-aligns+}. For the local structure alignment, due to the lack of appropriate datasets and quantitative evaluation metrics, we investigate the model's performance through a case study. All experiments were conducted on 8 A800 GPUs, each with 80GB VRAM. The implementation will be released upon acceptance.

\subsection{\textbf{CATH-aligns}: Benchmark for Structure Alignment}
We construct \textbf{CATH-aligns}, a new benchmark with standard quantitative evaluation criteria. We process the dataset from \textbf{CATH 4.3} \footnote{Official dataset can be found at \url{http://download.cathdb.info/cath/releases/all-releases/v4\_3\_0/}}, a comprehensive dataset with experimentally determined protein domain structures. All structures are labeled with a four-level CATH classification code \cite{ORENGO19971093cath} that classifies the protein's structural type from different perspectives. We remove incomplete protein entities that include missing atomic coordinates for $C_{\alpha}$ and $N$. All proteins are below $20\%$ of sequence identity to each other. A total of $14,654$ are left for constructing the independent test set \textbf{CATH-aligns}.


For structure alignment prediction, we define a binary classification task with the split test subset from \textbf{CATH4.3}. We consider two levels of classification difficulty and name them as \textbf{CATH-aligns} and \textbf{CATH-aligns+}, respectively. The former \textbf{CATH-aligns} defines negative pairs as protein domains with all the four-level CATH classification codes being different and positive pairs as any of the four codes being identical. The latter \textbf{CATH-aligns+} defines a more difficult task, where structure pairs with identical CATH codes at all four levels are considered positive sample pairs, while pairs differing at any level are considered negative sample pairs. To ensure computational efficiency and balance the number of positive and negative samples, we prepare three folds for evaluation, each containing $10,000$ positive and $10,000$ negative pairs that are randomly sampled from the complete $14654\times 14654$ pairs of \textbf{CATH-aligns}. The prediction results are assessed using the AUC (area under the curve) metric, where an AUC closer to $1$ indicates better predictive performance.

\subsection{Experimental Protocol}


\paragraph{Training Setup}
\loca~is optimized with \textsc{AdamW} \cite{kingma2014adam} with a learning rate of $0.0001$. The maximum number of training epochs is set to $50$, and early stopping is applied with a patience of $5$ epochs. For stable memory usage of GPU during the training, the maximum number of nodes per batch is set to $10,000$. The GVP module consists of $6$ layers and a dropout ratio of $0.2$. The embedding dimensions are set to $256$ for $\mN_S$, $32$ for $\mN_V$, $64$ for $\mE_S$, and $2$ for $\mE_V$. During the inference, the input $\mN_S$ is masked to $0$ to obtain the representation vectors for each point in the protein structure. All experiments are conducted on an A800 GPU with 80GB of memory, and the training process is logged using WanDB.

\paragraph{Dataset for Self-Supervised Learning}
We use unlabeled \textbf{CATH4.3\_s40} for training our graph representation learning model. All structures in the dataset are processed with the similarity threshold at $40\%$, containing a total of $31,070$ protein domain structures. The training target is to recover the noisy input tokens defined in Section~\ref{sec:global_train_objective}. A subset of $200$ domains is split randomly for model validation. Although the training dataset is unlabeled, we further ensure that the sequence identity between the training set and the test datasets \textbf{CATH-aligns} is below $20\%$ to avoid data leakage.



\begin{figure*}[th]
    \centering
    \includegraphics[width=0.9\textwidth]{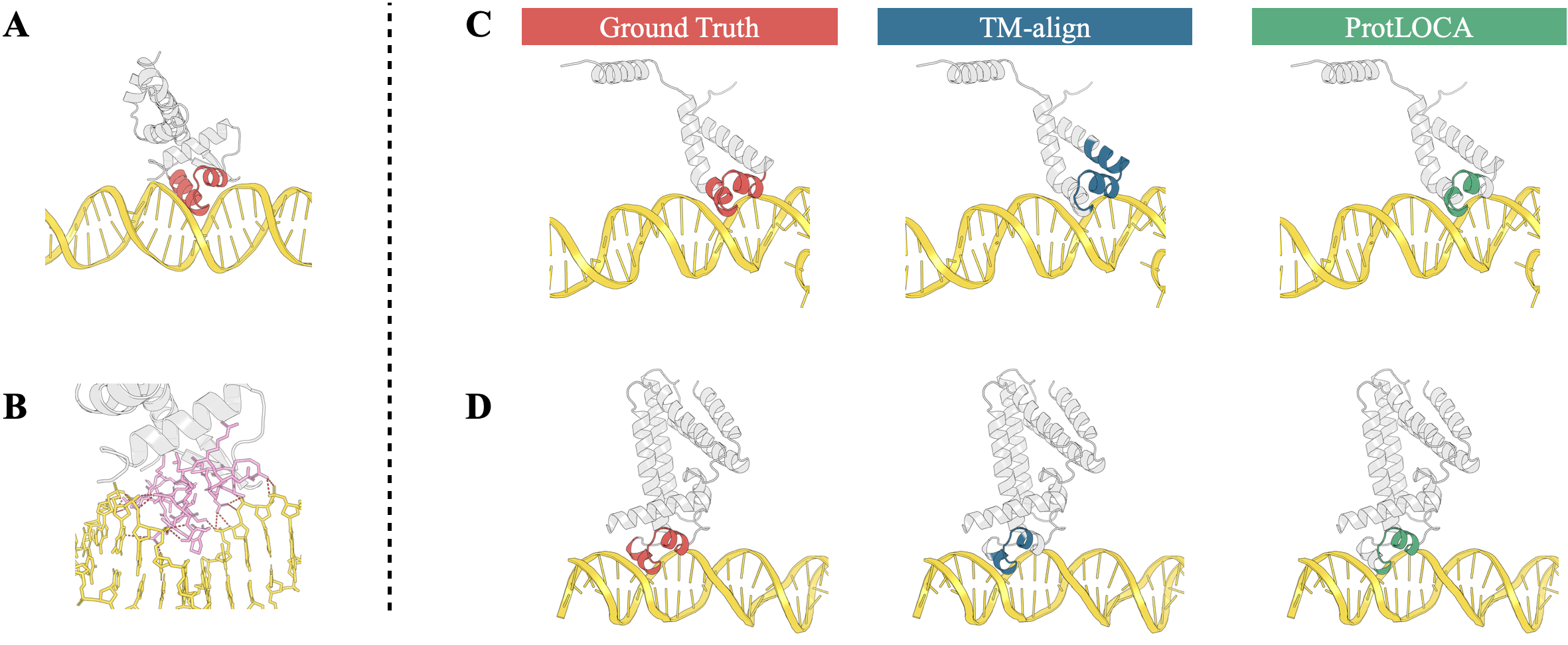}
    \caption{Example of using \loca~and TM-align to find Helix-turn-helix (HTH) motif in DNA binding protein. (A) HTH motif in Tox repressor (PDB: 1F5T). The HTH motif is colored in red, DNA in yellow, and protein in white. (B) The HTH motif serves as the binding site of protein to DNA and is presented as a Tox repressor. The HTH motif is colored in pink, the protein is in white, the DNA is in yellow, and the hydrogen bonds between the HTH motif and DNA are marked in red. (C) phage lambda cII protein (PDB: 1ZS4) HTH motif from ground truth (red), TM-align (blue), and \loca~(green). (D) transcriptional regulator PA2196 (PDB: 4L62) HTH motif from ground truth (red), TM-align (blue), and \loca~(green).}
    \label{fig:caseStudy_HTH}
\end{figure*}

\paragraph{Baseline Methods}
We compare \loca~with a set of alignment-based and embedding-based deep learning methods. For alignment methods, we consider two variants of FoldSeek \cite{van2024fast}, using 3Di with pure structural input and 3Di with both structural and amino acid (AA) input. This method encodes local structures and uses traditional alignment algorithms for point-by-point comparison of structures. 
For the global structure matching task, we exclude TM-align \cite{zhang2005tm} from the baseline list due to its extremely inefficient computational speed. In order to compute the similarity of all $14654\times 14654$ structure pairs in the test dataset, TM-align would consume approximately $30,000$ hours. In comparison, \loca~spends less than $1$ hour, including the data preprocessing and scoring steps. 
For embedding methods, our comparison includes the pre-trained sequence-based language model ESM2 \cite{lin2023esm2} with different model scales. The structure-aware pre-trained model ProstT5 \cite{heinzinger2023prostt5} uses both AA2fold and fold2AA modes for translation tasks, we take amino acid sequences and Foldseek sequences as input to get embeddings respectively. We also include two inverse-folding methods, ESM-if1 \cite{hsu2022esmif} and MIF-ST \cite{yang2022mifst} which take amino acid sequences as input. Unlike alignment methods, embedding methods average protein sequences to obtain embeddings and use the dot product of these vectors to measure overall protein similarity.

\subsection{Results Analysis}
\paragraph{Baseline Comparison}
Table~\ref{tab:result_main} reports the performance comparison of \loca~and other baseline models on \textbf{CATH-aligns} and \textbf{CATH-aligns+}. In both alignment tasks, \loca~significantly outperforms other embedding methods and even exceeds the performance of the classic alignment-based baseline FoldSeek. Note that the training cost for \loca~is lower than that of all baseline methods due to a significantly smaller number of trainable parameters. Additionally, it is trained on a considerably small dataset of approximately $30,000$ samples. This training set size is smaller than what is typically required for deep protein models, which usually demand millions or more samples to train effectively. Furthermore, structure-based algorithms (\eg ESM-if1) generally perform better than sequence-based methods. Notably, ESM2, despite achieving state-of-the-art performance in many downstream tasks, does not perform well in the structure alignment task. Additionally, the results of both FoldSeek and \loca~demonstrate that incorporating amino acid information during training can indeed reduce the overall predictive performance of the models. These experimental results strongly support our initial claim that amino acid information does not always contribute to learning more expressive hidden embeddings, and embeddings learned with sequence information do not consistently enhance the prediction performance in any downstream tasks.

\paragraph{Sensitivity Analysis}
We examine the impact of two hyperparameters on the performance of \loca: the masking noise ratio $p$ (with the permutation ratio being $1-p$) and the number of GVP layers. The results are visualized in the left two subplots in Fig.~\ref{fig:sensitivity}. The prediction accuracy is insensitive to both hyperparameters, with less than $1\%$ changes observed from a considerably large range. We perform $p=0.5$ and $6$ GVP layers as the default settings for the model.

\paragraph{Input and Denoising Token}
Fig.~\ref{fig:sensitivity} (right) compares the effect of different types of input node features. We consider three types of node features: the classic amino acid type, the secondary structure codes (DSSP), and the hidden structure codes (3Di). Overall, using 3Di encoding yields the best prediction performance on the structure alignment task. More importantly, incorporating amino acid information during the model training significantly degrades model performance (green bars). This observation is consistent with the previous analysis and our key assumption, where considering amino acid information in the structure alignment task may introduce unnecessary interference, leading to poor prediction performance in downstream tasks.

\subsection{Case Study: HTH Functional Structure Alignment}
The helix-turn-helix (HTH) motif is a crucial structural component in DNA binding proteins, including transcription factors regulating gene expression \cite{takeda1983dna}. It comprises two alpha-helices joined by a `turn', with the second helix, known as the recognition helix (Fig.~\ref{fig:caseStudy_HTH}A), specifically interacting with DNA (Fig.~\ref{fig:caseStudy_HTH}B). This interaction is essential for gene regulation, as it enables proteins containing the HTH motif to control the transcription process by attaching to DNA's promoters or operators \cite{ishihama2010prokaryotic}. We first use TM-align to identify the HTH motif in the phage lambda cII protein (Fig.~\ref{fig:caseStudy_HTH}C) \cite{jain2005crystal} and the transcriptional regulator PA2196 (Fig.~\ref{fig:caseStudy_HTH}D) \cite{kim2013crystal}. In the phage lambda cII protein, the HTH motif identified by TM-align is located differently in protein compared to its position in the ground truth. For transcriptional regulator PA2196, the HTH motif identified by TM-align is much shorter than the one in the ground truth. These cases demonstrate  TM-align's limitations in accurately identifying the correct HTH motif in DNA binding proteins. However, \loca~can effectively identify the correct HTH motif in these two proteins, despite their different overall folds. Thus, \loca~demonstrates better performance than TM-align in identifying critical motifs in proteins with the same functions when their overall structures vary.

\section{Related Work}
\subsection{Sequence Representation}
With the growth of protein sequences and advancements in natural language modeling methods, the most commonly used approaches in protein representation learning typically involve unsupervised training on protein sequences, without considering protein structural information. For example, ESM2 \cite{lin2023esm2}, ESM-1v \cite{meier2021esm1v}, and ESM-1b \cite{rives2021esm1b} use different redundancy levels of the Uniref dataset \cite{suzek2015uniref}, employing the BERT \cite{devlin2018bert} architecture and a masked language modeling unsupervised training objective to train models for downstream tasks related to representation learning or zero-shot mutation tasks in protein engineering. ProtTrans \cite{ahmed2021prottrans} has introduced a series of protein language representation models, such as ProtBert, ProtT5, and ProtAlBert, based on BERT \cite{devlin2018bert}, T5 \cite{raffel2020t5} or AlBert \cite{lan2019albert} architectures, primarily applied to various downstream tasks of representation learning. Ankh \cite{elnaggar2023ankh} uses an asymmetric encoder-decoder approach and explores a series of training parameters to train language models that perform well on downstream tasks. Additionally, methods like CARP \cite{yang2022carp} and ProteinBert \cite{brandes2022proteinbert} use 1D-CNN instead of the attention mechanism to improve training efficiency for processing longer sequences. 

\subsection{Structure Representation}

With the increase in crystal structures and advancements in folding techniques \cite{jumper2021alphafold, lin2023esm2}, protein structure databases have become increasingly large \cite{varadi2022afdb}. Currently, mainstream methods use sequence information as the training target for structural inputs or as auxiliary node features, with few models considering only protein structures while discarding amino acid types. For instance, GearNet \cite{zhang2022gearnet} uses contrastive learning to enhance representation quality for protein enzyme commission (EC) number prediction, and ProtLGN \cite{zhou2023protlgn} employs multi-task learning and denoising training objectives to improve zero-shot prediction capabilities for protein mutations. Additionally, models like GVP \cite{jing2020gvp} and EGNN \cite{satorras2021egnn} use graph neural networks to model the equivariance and invariance of proteins for protein quality prediction tasks. Some inverse folding methods use protein structures as input to restore amino acid information, achieving structure-aware training. For example, ESM-IF \cite{hsu2022esmif} uses GVP to initialize transformer node features, and ProstT5 \cite{heinzinger2023prostt5} uses Foldseek's structural tokens as input and amino acid sequences as output (or vice versa) for machine translation training. Furthermore, some approaches combine language models and graph neural networks to enhance the quality of representation learning. Examples include MIF-ST \cite{yang2022mifst}, which integrates CARP \cite{yang2022carp} and Struct GNN, ProtSSN \cite{tan2023protssn}, which combines ESM2-650M and EGNN structures, and LM-GVP \cite{wang2022lm-gvp}.


\section{Conclusion and Discussion}
Protein function annotation and analysis typically rely on protein sequences and overall structural information. However, these approaches come with their own set of challenges. Sequence-based analysis, such as EC numbers and Pfam datasets, doesn't consistently yield accurate analysis. This is partly because pinning down a protein’s position on the evolutionary tree can be problematic when only its sequence is considered. In addition, methods that align overall protein structures, such as TM-align, may overlook proteins that are characterized by local structural conservation while amidst overall structural variability. Protein functions are mainly determined by key sub-structures, such as catalytic region and binding pockets, while the remaining structures determine the physical properties of proteins. In light of these issues of current methodologies and the significance of biology, we developed the \loca~that focuses on local structural matches within proteins with diverse overall folds. This tool unlocks new perspectives on protein functional and structural evolution.

\bibliographystyle{IEEEtran}
\bibliography{1reference.bib}

\clearpage

\end{document}